\documentclass[journal=jacsat,manuscript=article]{achemso}

\usepackage{subcaption}
\usepackage{placeins}
\usepackage{amsmath}

\author{Nic\'{e}phore Bonnet}
\email{nicephore.bonnet@epfl.ch}
\author{Nicola Marzari}
\affiliation{Theory and Simulation of Materials (THEOS), Ecole Polytechnique F\'{e}d\'{e}rale de Lausanne, 1015 Lausanne, Switzerland}

\title[An \textsf{achemso} demo]
  {Ion Sieving in Two-Dimensional Membranes from First Principles}

\abbreviations{IR,NMR,UV}
\keywords{American Chemical Society, \LaTeX}

\begin{document}

\begin{abstract}
A first-principles approach for calculating ion separation in solution through two-dimensional (2D) membranes is proposed and applied. Ionic energy profiles across the membrane are obtained first, where solvation effects are simulated explicitly with machine-learning molecular dynamics, electrostatic corrections are applied to remove finite-size capacitive effects, and a mean-field treatment of the charging of the electrochemical double layer is used. Entropic contributions are assessed analytically and validated against thermodynamic integration. Ionic separations are then inferred through a microkinetic model of the filtration process, accounting for steady-state charge separation effects across the membrane. The approach is applied to Li$^{+}$, Na$^{+}$, K$^{+}$ sieving through a crown-ether functionalized graphene membrane, with a case study of the mechanisms for a highly selective and efficient extraction of lithium from aqueous solutions.   

\end{abstract}

Keywords: 2D membranes, ion sieving, first-principles calculations, machine learning, microkinetic model, multiscale modelling, electrochemical double layer

\section{Introduction}

Advanced membrane filtration is an important target for emerging separation technologies, and is critically needed, e.g., for sustainable water and energy management \cite{Shao2020,DuChanois2021}. Notably, ion selectivity is an increasingly required feature for membrane technologies\cite{Werber2016}. In water treatment and recycling, ion-selective membranes can be used for recovery of valuable resources, such as lithium from brines, rare-earth elements from industrial wastewater, or phosphate and nitrate from wastewater, or for targeted removal of undesired species, such as heavy metals \cite{DuChanois2021,Stringfellow2021}. Ion selectivity is equally important for advanced membranes used as half-cell separators in electrochemical technologies for low-carbon energy conversion and storage, including batteries, fuel cells, and electrolyzers. Current polymeric membranes have been unable to reach the desired separation target for many advanced water and energy applications owing to the permeability-selectivity trade-off, whereby higher selectivities entail smaller permeabilities. This trade-off has been linked to the inability to control satisfactorily the pore size distribution in polymeric membranes, which motivated the development of novel materials with a higher molecular-level control over physical and chemical properties; these include porous crystalline materials, 2D materials, and biomimetic materials \cite{DuChanois2021,Mounet2018,Zhao2024}.     

In parallel, optimal design and operation of ion-sieving membranes requires a more fundamental understanding of the molecular drivers for ion selectivity; notably, the interplay between size, charge, dielectric and chemical effects \cite{Szymczyk2005,Yuan2022,Wen2016,Razmjou2019,Fan2023}. Size effects include the need for the ions to rearrange or remove their hydration shell to fit within the pores, and are thus related to both solvation and dielectric effects. Charge effects occur by intrinsic charging of the membrane and/or by charge separation across the membrane, e.g. via the Donnan mechanism \cite{Fievet2014}. Finally, pores' functionalizations may bring dramatic selectivity effects through chemical coordination, as exemplified by crown-ether graphene membranes \cite{Fang2019,Sahu2019,Guo2021,Lv2023}.

While data-driven approaches have been successfully applied to rank the importance of selectivity drivers in existing polymeric membranes \cite{Ritt2022}, atomistic simulations are needed for a detailed mechanistic understanding and computational screening of new candidate ion-selective materials \cite{DuChanois2021,Zhou2020,Chipot2023}. Classical molecular-dynamics (MD) simulations based on traditional force fields, which were first applied to investigate the permeability and selectivity of membranes \cite{Fang2019,Sahu2019,Sint2008,Suk2010,Ruan2016,Smolyanitsky2018,Zofchak2022,Meng2023}, can be limited by the accuracy of the models. In first-principles simulations, alternatively, an explicit treatment of all solvent molecules may be computationally burdensome, and implicit solvent models \cite{Andreussi2012}, while useful and able to reduce the computational costs, may fail to capture specific molecular effects at the solute / solvent interface. Moreover, charge effects are embedded within the electrochemical double layer (EDL), a complex structure surrounding the membrane and often exceeding the typical size of computational cells \cite{Shin2022}. Finally, the thermodynamic potential energy surfaces of ionic species across the membrane translate into effective ionic selectivities only in the context of a dynamic description of the membrane filtration process \cite{Szymczyk2005}.

The present work proposes a first-principles-based methodology for the prediction of membrane ion separation and applies it to the case of a paradigmatic 2D advanced membrane, addressing all the previous points. The energy profiles (EPs) of individual ions across the membrane pore are first determined, where solvation effects are simulated explicitly by machine-learning (ML) accelerated MD, and electrostatic correction schemes are applied to remove capacitive effects arising in finite-size simulation cells. EDL effects are added subsequently as a mean-field contribution to the ionic EP, using simplifying assumptions on the structure of the EDL above certain ionic concentrations. Entropic effects are assessed analytically and through a thermodynamic integration scheme. Finally, a microkinetic model of the ion-sieving process is developed to obtain effective ionic separations for a realistic set of operational parameters of the dynamic environment. The methodology is illustrated with the case of Li$^{+}$, Na$^{+}$, K$^{+}$ separation through 12-crown-4 ethers embedded in a graphene membrane in aqueous solution.    
                       
\section{Results and Discussion}
\subsection{Methodological overview}
\begin{figure}\begin{center}
\includegraphics[width=0.8\columnwidth]{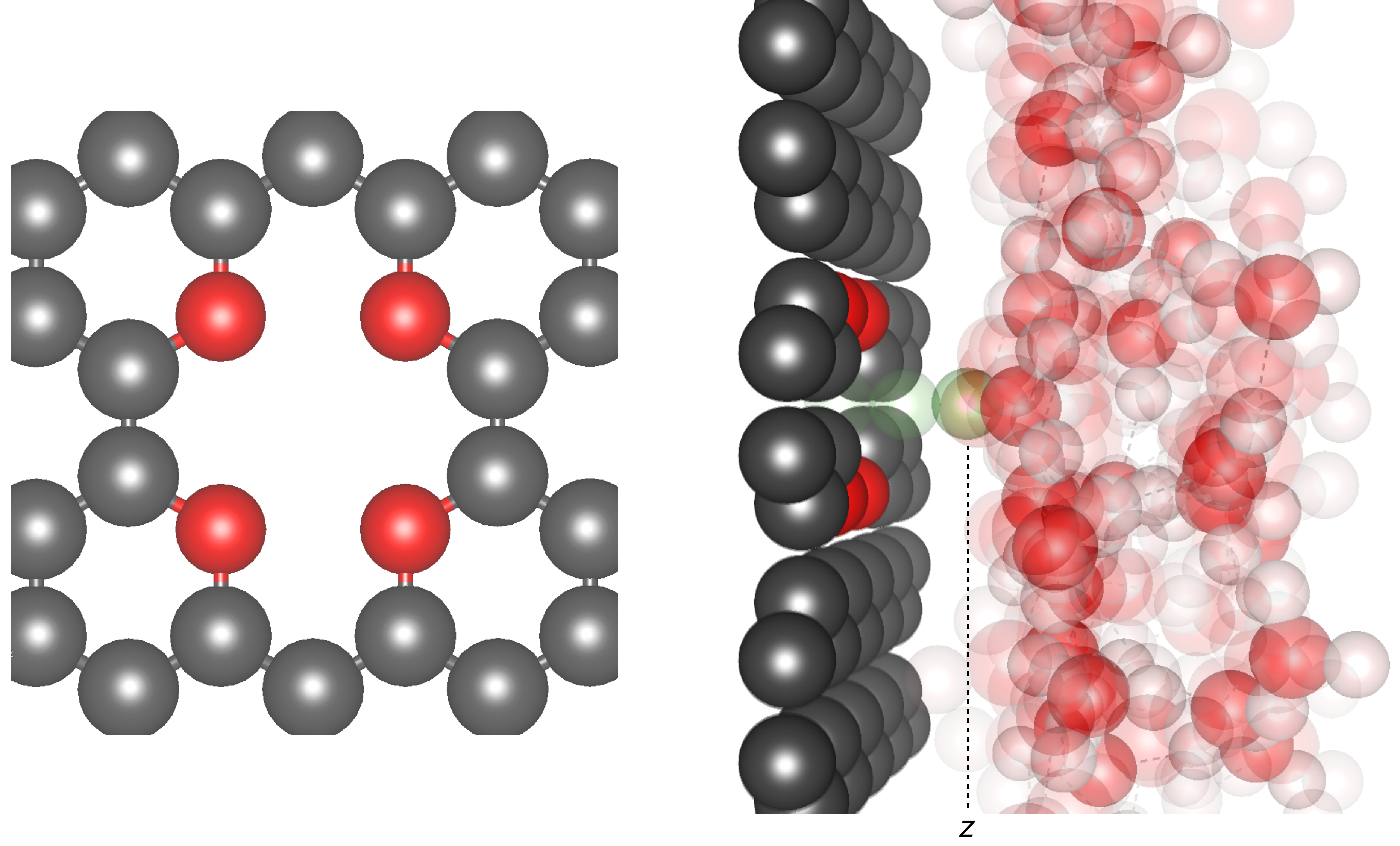}
\caption{Top (left) and side (right) view of the membrane pore and of the system setup to calculate the ion translocation energy profile; here for the case of a Li$^{+}$ ion solvated in water across a crown-ether functionalized graphene membrane.}
\label{top_side_view}
\end{center}\end{figure}

The translocation of an ion through a membrane is considered via the setup of Fig. \ref{top_side_view}. Unless stated otherwise, $z$ ({\AA}) will denote the longitudinal distance of the ion to the membrane plane. For a small pore with an activated translocation, as considered in the present study, the ion permeance is mainly governed by a two-step process \cite{Yuan2022}: first, the formation of a pore-associated state of the permeating ion, typically as an adsorption state in the pore vicinity; second, the actual translocation event from the pore-associated state. To quantify this process, we determine the energy profile (EP) of the ion in the pore region as a function of $z$. The solvated EP is decomposed as 
\begin{equation}
E_{aq}(z) = E_{vac}(z) + E_{solv}(z)
\end{equation}
where $E_{vac}(z)$ is the EP in vacuum and $E_{solv}(z)$ is the solvation energy. In the following, we present an approach to calculate $E_{vac}(z)$ and $E_{solv}(z)$ from first principles detailing the intermediate steps using the example of a solvated Li$^{+}$ ion, as in Fig. \ref{top_side_view}. The simulation cell contains one pore made of a 12-crown-4 ether analogue embedded in the graphene membrane. Two cell sizes are considered: a first simulation cell, denoted as $1\times 1$, with dimensions 12.33 {\AA} $\times$ 12.81 {\AA}; and a larger simulation cell, denoted as $2 \times 2$, with dimensions 24.66 {\AA} $\times$ 25.62 {\AA}. Total energy calculations are performed in vacuum or in the presence of explicit water.

Entropic contributions are assessed analytically and via thermodynamic integration, and the resulting free energy profile is injected into a microkinetic model of the sieving process. The model incorporates EDL effects as a mean-field contribution to ionic EPs and self-consistently predicts the charge separation density across the membrane upon filtration. 

\subsection{Energy profile in vacuum}

\begin{figure}\begin{center}
\includegraphics[width=0.95\columnwidth]{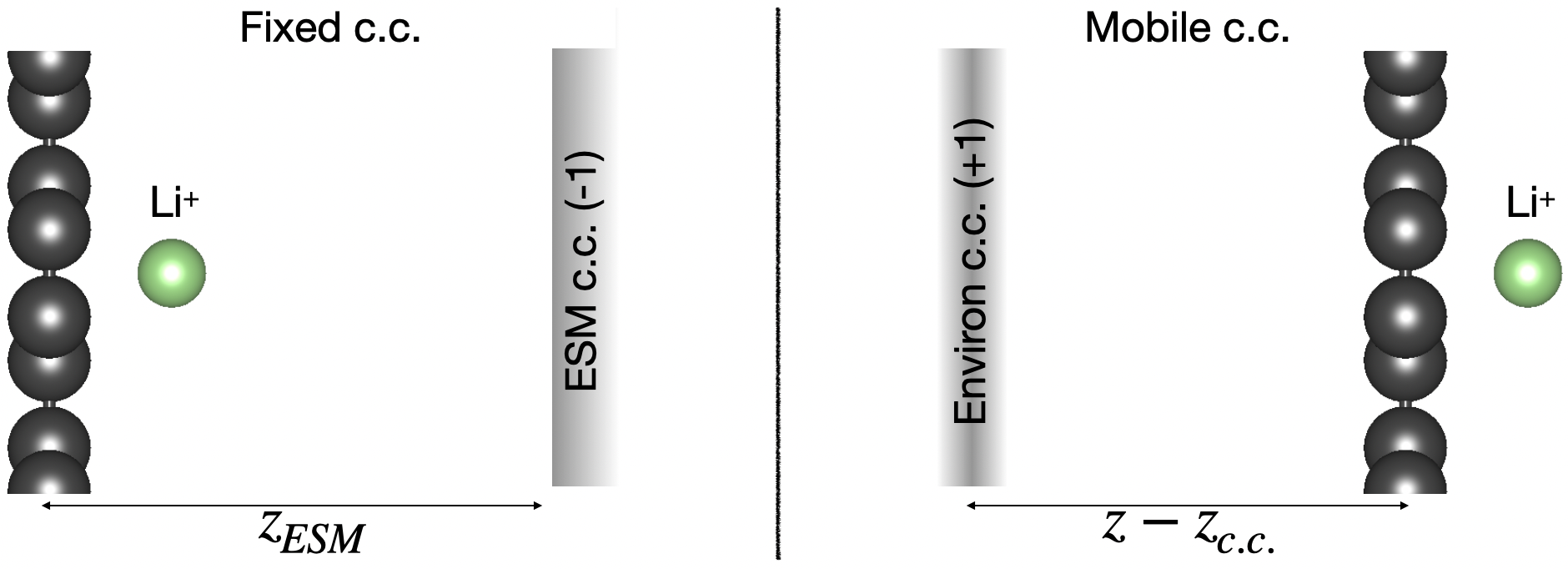}
\caption{Countercharge (c.c.) setups. Left panel: fixed c.c. applied with the effective screening medium (ESM) counterelectrode\cite{Otani2006} ($z_{ESM} = $ constant). Right panel: mobile c.c. applied using the self-consistent continuum solvation model\cite{Andreussi2012}, as encoded in the Environ library\cite{Environ}, with $z-z_{c.c.} =$ constant.}
\label{countercharges}
\end{center}\end{figure}

The EP in vacuum is first addressed. In principle, $E_{vac}(z)$ is obtained as the Boltzmann-weighted mean energy of the ion at different transverse positions above the pore. As a simplification, it is here taken as the energy of the pore-centered position in the 1$\times$1 cell, and as the minimum energy in the 2$\times$2 cell, at fixed $z$. The translational multiplicity of the pore-associated state in the transverse plane is at most $W \approx \frac{A_{pore}}{\lambda_{d}^{2}}$, where $A_{pore}$ is the pore surface area, $\lambda_{d} = \frac{h}{\sqrt{2\pi m_{Li} kT}}$ is de Broglie's thermal wavelength, $k$ is Boltzmann's constant, $h$ is Planck's constant, and $m_{Li}$ is the mass of the lithium ion\cite{Chorkendorff2007}. From Boltzmann's definition of entropy, the absolute difference between the minimum energy and the free energy is thus at most $kT ln(W) \approx 0.1$ eV for the present pore ($A_{pore} \approx 7$ \AA$^{2}$) at room temperature. 

The total energy is calculated with density-functional theory (see Methods), where the explicit system bears the net +1 charge of the cation. Implicit countercharges (c.c.) are introduced for technical reasons which will be clarified below. Three different c.c. setups are considered: (i) no c.c. added; (ii) a fixed neutralizing c.c. ($-1$) mimicking a Stern-type c.c.\cite{Shin2022} using the Effective Screening Medium (ESM) counterelectrode of Otani and Sugino \cite{Otani2006}, placed on the right-hand side of the cell (Fig. \ref{countercharges}, left panel); and (iii) a mobile 2D Gaussian c.c. applied with the self-consistent continuum solvation\cite{Andreussi2012} using the Environ solvation library \cite{Environ}, keeping the ion-c.c. distance ($z - z_{c.c.}$) constant (Fig. \ref{countercharges}, right panel). For convenience, the ``countercharge" of case (iii) is artificially placed on the opposite side of the membrane and is thus taken positive for symmetry reasons. In all cases, calculations are performed within periodic-boundary conditions (PBC), but an electrostatic correction \cite{Dabo2008} is applied to impose open-boundary conditions (OBC) -- that is, to remove the effect of periodic images -- in the longitudinal $z$ direction. Moreover, charge separations give rise to capacitive artefacts in the finite simulation cell \cite{Chan2015}; in the following, the EP of setup (ii) is thus obtained by applying a capacitive correction (see Methods) to the calculated total energy. 

\begin{figure}
\centering
\begin{subfigure}{.5\textwidth}
  \centering
  \includegraphics[width=.97\linewidth]{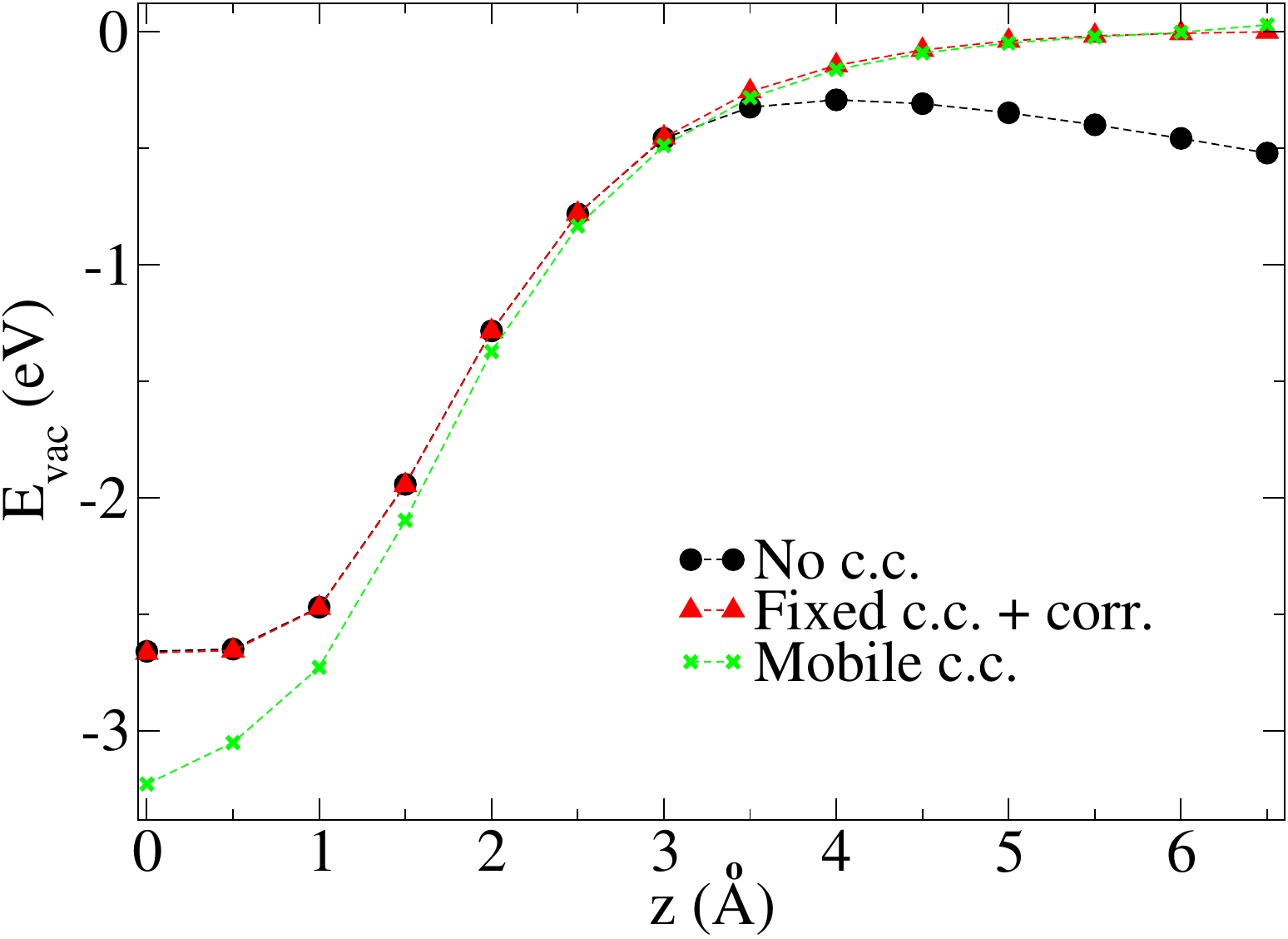}
\end{subfigure}%
\begin{subfigure}{.5\textwidth}
  \centering
  \includegraphics[width=.97\linewidth]{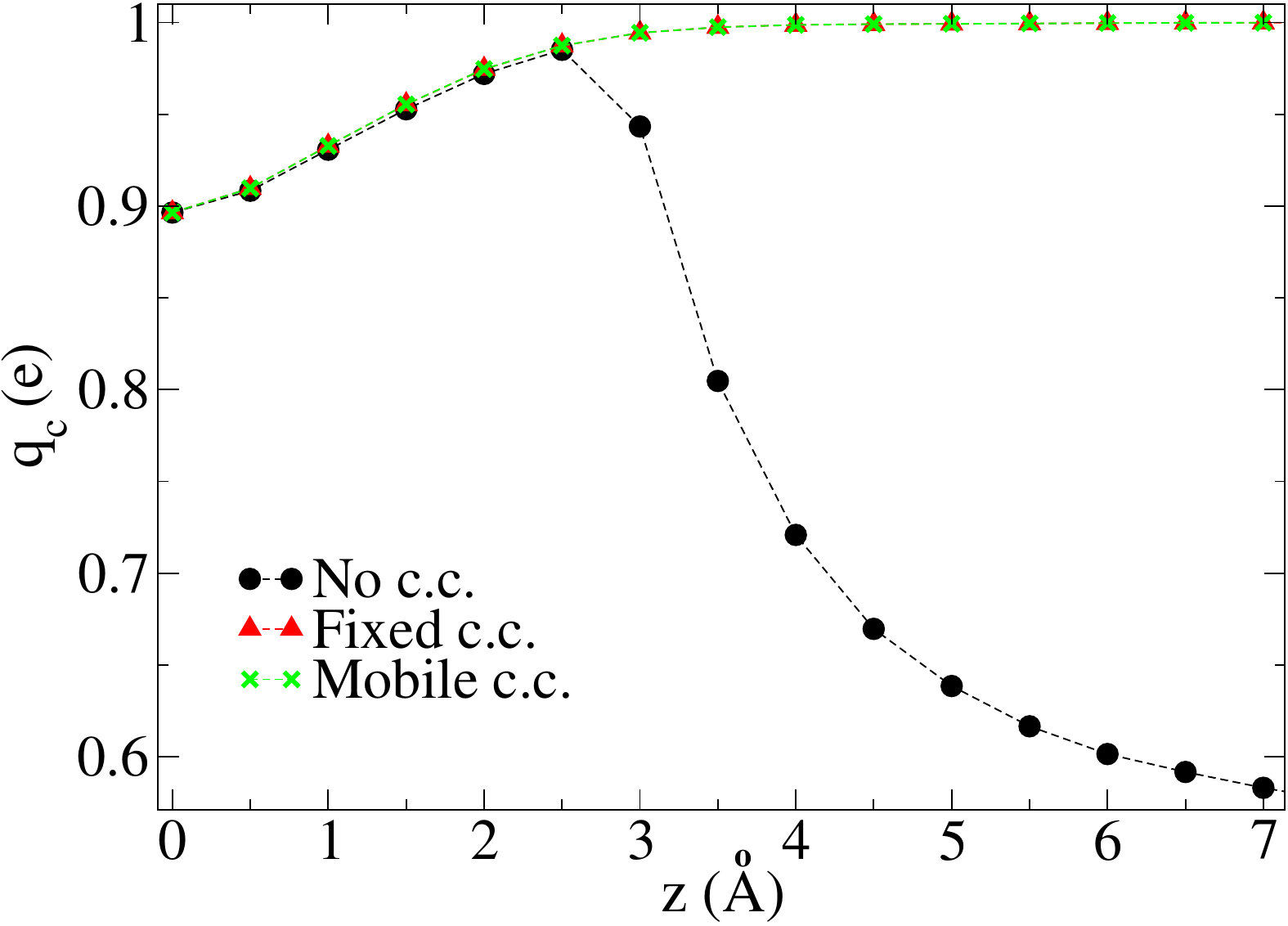}
\end{subfigure}
\caption{Left panel: $E_{vac}(z)$ profiles for lithium through the unrelaxed membrane in the $1\times 1$ cell with the different c.c. setups. Right panel : The corresponding net charge $q_{c}(z)$ on lithium.}
\label{energies_charges_33}
\end{figure}

Fig. \ref{energies_charges_33} shows the energy profile $E_{vac}(z)$ and net charge of the cation $q_{c}(z)$ for lithium going through the unrelaxed membrane in the $1\times 1$ simulation cell and within the different c.c. setups, where the charge $q_{c}(z)$ is obtained by a Bader analysis\cite{Tang2009}. The following behaviours can be noted:
\begin{itemize}
\item At short distances, the $q_{c}(z)$ curves are identical and lower than 1 due to chemical interactions with the membrane, and converge to 1 (as expected for the isolated Li$^{+}$ cation) for $ z $ approaching 3 {\AA} and beyond. However, in the absence of a c.c., $q_{c}(z)$ decreases again for $z>3$ {\AA}. Electronic charge spilling from the membrane is responsible for this behaviour and is suppressed by adding the c.c., as the additional electric field generated tends to destabilize extra electrons on Li$^{+}$. For some systems, the absolute value of the ``countercharge" required to prevent the electronic spillover may be larger than 1 (up to 3 in the present work), in which case the ESM setup is not applicable and only the Environ setup can be used.

\item For $z>3$, the $E_{vac}(z)$ curves obtained from the two c.c. methods are similar, as for these ion positions the two methods are electrostatically equivalent. However, for $z<3$, energies of the mobile c.c. setup are artificially reduced by an uncorrected finite-size effect, as for these ion positions the displacement field created by Li$^{+}$ and its c.c. is partially screened by the membrane, in contrast with other ion positions.

\item For $z<3$, the $E_{vac}(z)$ curves without a c.c. and with the fixed c.c. are similar, as expected from the capacitive correction. However, they become different for $z>3$ owing to their different charges on Li$^{+}$.   
\end{itemize} 

Overall, the $E_{vac}(z)$ curve can thus be obtained, with the correct charge on the ion and without finite-size capacitive energies, by the following approach: at large distances ($z>3$), the membrane charge spilling is prevented by applying a sufficiently large c.c., and energies are obtained with the mobile c.c. setup; at short distances ($z < 3$), the energies are obtained without the c.c. In both cases, the corrected fixed c.c. setup (when applicable) gives similar results. This approach is used to obtain $E^{1}_{vac}(z)$ in the 1$\times$1 simulation cell with the membrane unrelaxed, and $E_{vac}^{2}(z)$ in the $2\times 2$ simulation cell with the membrane relaxed.  

\subsection{Energy profile in water}
Next, the EP in water is determined. In the $1\times 1$ simulation cell, $E_{aq}^{1}(z)$ is calculated first as the mean total energy $\langle E(z,\nu) \rangle_{\nu}$ of the system over configurations $\nu$ of the explicit solvent as sampled by machine-learning molecular dynamics (see Methods). The solvation energy is then inferred as $E_{solv}(z) = E_{aq}^{1}(z) - E_{vac}^{1}(z)$. Alternatively, to include entropic effects, the solvation free energy profile $G_{solv}(z)$ can be obtained directly and exactly using thermodynamic integration \cite{Milman1993}: 
\begin{equation}
G_{solv}(z) = \int_{z}^{\infty} \left( \langle F_{aq}^{Li}(z,\nu) \rangle_{\nu} - F_{vac}^{Li}(z) \right)dz
\label{eq:ti}
\end{equation}
where $F_{vac}^{Li}(z)$ is the longitudinal force on Li$^{+}$ in vacuum, and $\langle F_{aq}^{Li}(z,\nu) \rangle_{\nu}$ is the mean longitudinal force on Li$^{+}$ in water, as calculated from the MD trajectories. Finally, the solvated EP is inferred in the 2$\times$2 cell as $E_{aq}^{2}(z) = E_{vac}^{2}(z) + E_{solv}(z)$.

\begin{figure}\begin{center}
\includegraphics[width=0.8\columnwidth]{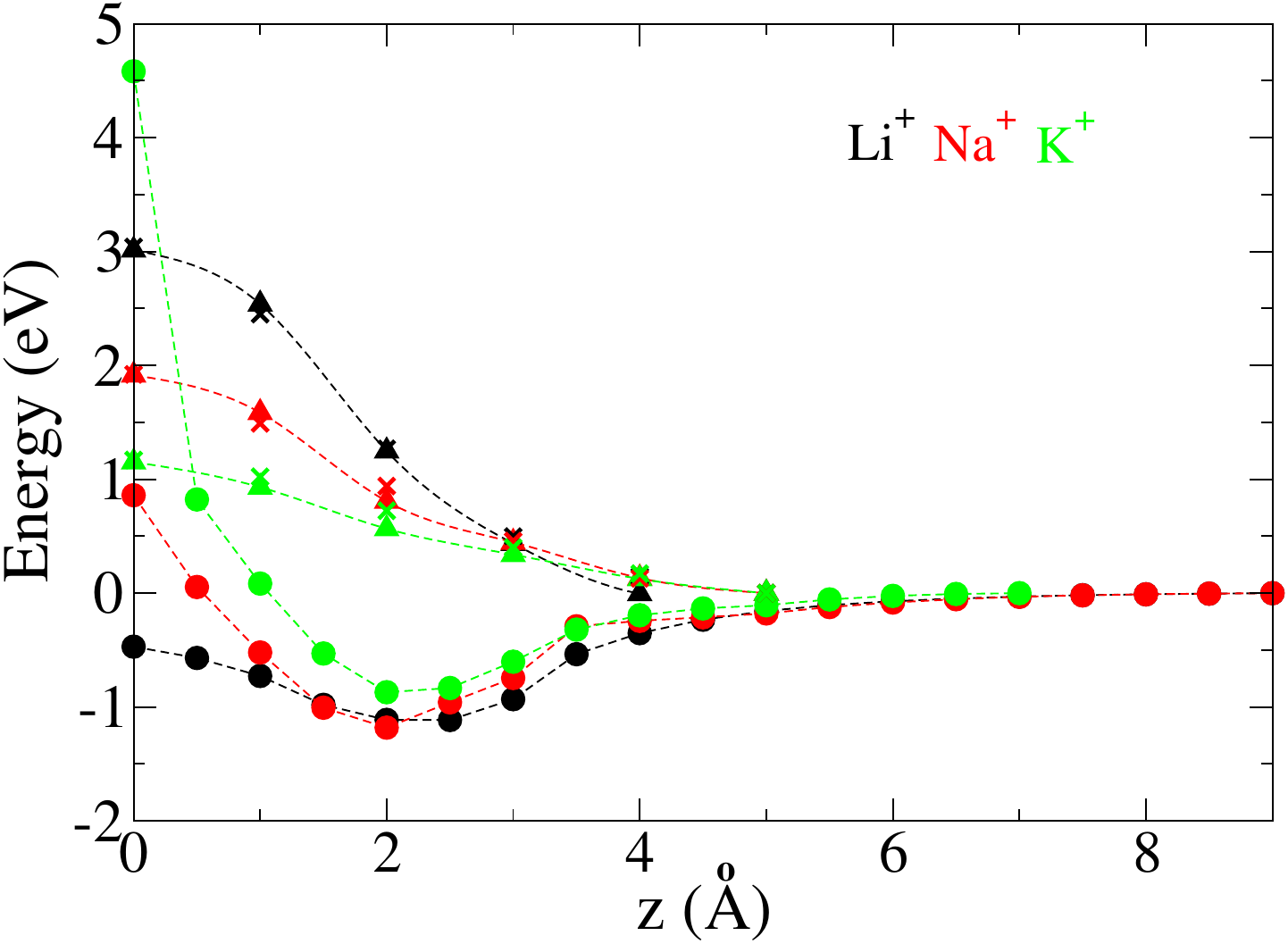}
\caption{Solvated energy profile $E^{2}_{aq}(z)$ (circles) and solvation energy contribution $E_{solv}(z)$ (triangles) for Li$^{+}$, Na$^{+}$, and K$^{+}$ through the membrane. The crosses indicate the solvation free energy $G_{solv}(z)$, as obtained from thermodynamic integration (Eq. \ref{eq:ti}).}
\label{all_pes}
\end{center}\end{figure}

Results are shown for Li$^{+}$, Na$^{+}$ and K$^{+}$ in Fig. \ref{all_pes}. Solvation effects appear to have a strong contribution in the EP as the ion needs to lose its inner solvation shell to cross the membrane. The desolvation penalty follows the order Li$^{+}$ $>$ Na$^{+}$ $>$ K$^{+}$, in line with their sizes and bulk hydration energies\cite{Conway1978}. Interestingly, for the three ions, the net EP resulting from chemical interactions with the membrane pore and desolvation effects exhibits the same global minimum (referred to as the adsorption energy in the following) of $\approx-1$ eV at $\approx 2$ \AA$ $ away from the membrane (referred to as the adsorption site). Solvation entropic effects are small, as evidenced by the close agreement between $E_{solv}(z)$ and $G_{solv}(z)$. Thus, the EPs are used as an approximation to the free energies within the dynamical model of the membrane filtration process.

We note that, within PBC, the solvation energy is not obtained strictly for a single ion and thus requires further interpretation. In the present case, an electrostatic analysis shows that the PBC electrostatic field in the first solvation shell region of the ion (within 3 {\AA} for Li$^{+}$ and Na$^{+}$)\cite{Xi2022} is similar to that for a single ion, while the contribution of the rest of space to $E_{solv}(z)$ is constant within $\pm$0.1 eV for all $z$. Consequently, $E_{solv}(z)$ can here be interpreted as the solvation energy of a single ion whose first solvation shell is affected by translocation, while the outer electrostatic energy is constant. In reality, the EDL potential bias brings an additional outer contribution to the ion energy profile. In the mean-field approximation, this contribution is taken as $E_{DL}(z) = e\Phi_{DL}(z)$, where $\Phi_{DL}(z)$ is the mean electrostatic potential profile inside the EDL.

\subsection{Parametric prediction of Li$^{+}$ extraction from a brine}
A microkinetic model of the membrane filtration process (see Methods) is used to predict steady-state ionic concentrations upstream (retentate side) and downstream (permeate side) of the membrane. The model input parameters include the individual ionic adsorption energies $E^{ads}_{i}$ from the bulk solution to the adsorption site, and translocation energy barriers $E^{*}_{i}$ from the adsorption site. EDL contributions are added self-consistently as a function of surface charges, particularly the Donnan surface charge densities denoted as $-\delta/+\delta$ on the retentate/permeate sides, respectively.  

The model predictions are illustrated with the following case study: an influent solution of ($1-x$) M Na$^{+}$, $x$ M Li$^{+}$, 1 M Cl$^{-}$; a retentate-to-permeate flow rate ratio of 1:1; a pore surface density of 1 $\mu$mol/m$^2$; a permeate velocity of 10 LMH (L/m$^2$/h) of the order of standard reverse osmosis velocities. In line with the first-principles results, $E_{Li}^{ads}$ and $E_{Na}^{ads}$ are set to -1.0 eV, and $E_{Na}^{*}$ to 2.0 eV. By analogy, $E_{Cl}^{ads}$ is also set to -1.0 eV. In this example, Cl$^{-}$ is envisioned to pass through other pore types than the cations, or even through a separate, anion-selective membrane. The steady-state filtration performance is then determined parametrically as a function of $x$, $E_{Li}^{*}$ and $E_{Cl}^{*}$.

\begin{figure}
\centering
\begin{subfigure}{.5\textwidth}
  \centering
  \includegraphics[width=.97\linewidth]{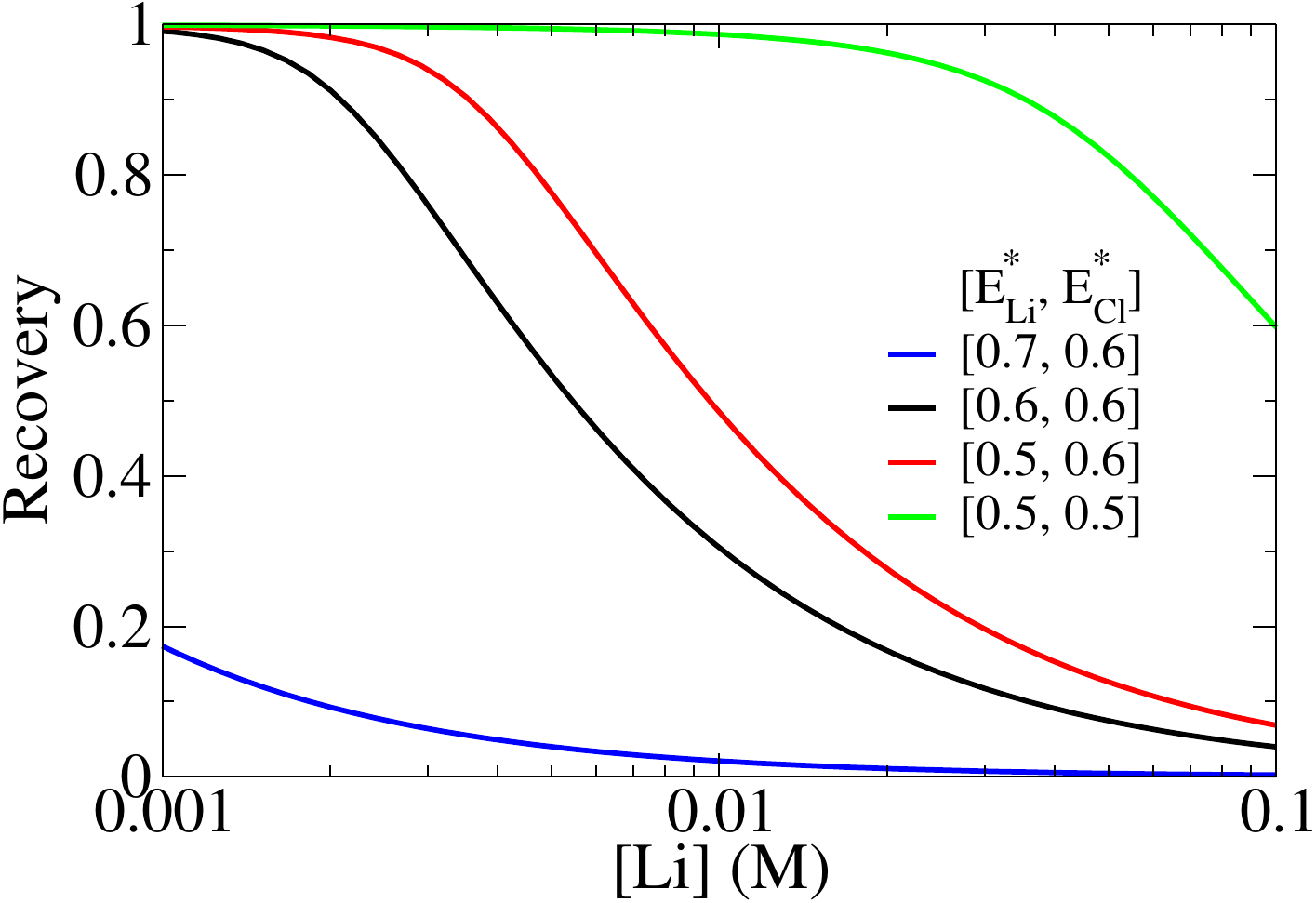}
\end{subfigure}%
\begin{subfigure}{.5\textwidth}
  \centering
  \includegraphics[width=.97\linewidth]{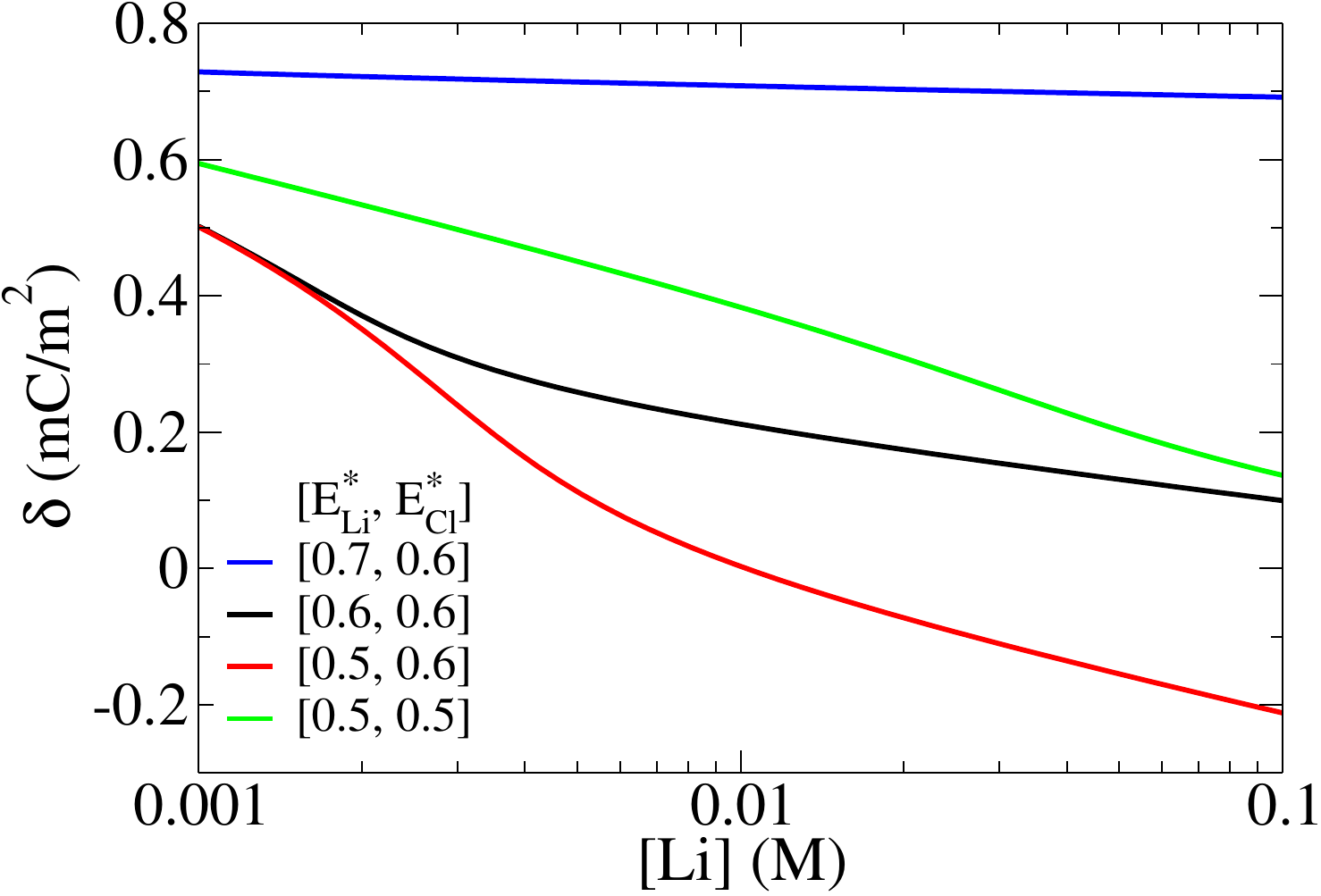}
\end{subfigure}
\caption{Lithium recovery in the permeate (left) and surface charge separation density (right) as a function of influent lithium concentration for different sets of $E_{Li}^{*}$ and $E_{Cl}^{*}$ values (eV). The first-principles value found for $E_{Li}^{*}$ through the 12-crown-4 ether graphene membrane is $\approx$ 0.6 eV (cf. Fig. \ref{all_pes}).}
\label{Li_results}
\end{figure}

Because of its high translocation barrier, Na$^{+}$ is essentially always blocked by the membrane, with a permeate concentration lower than 10$^{-8}$ M in all cases. By contrast, Fig. \ref{Li_results} shows the Li$^{+}$ concentration on the permeate side normalized to the maximum concentration of 2$x$ M, equivalent to a 100\% recovery of the influent lithium. As expected from the electroneutrality condition, the lithium recovery is affected by both $E_{Li}^{*}$ and $E_{Cl}^{*}$. Moreover, the charge density $\delta$ increases for smaller lithium concentrations as a mechanism to keep the lithium and chloride fluxes equal across the membrane, and thus promoting higher lithium recoveries. Accordingly, the highly selective (pure Li salt permeate) and efficient (up to 100\% lithium recovery) separation relies respectively on (i) the low translocation barrier of Li$^{+}$ relatively to other cations (ii) the ``entrainment" effect of the counter-ion (Cl$^{-}$) which, by the Donnan effect, electrostatically pulls lithium to the permeate side against the diffusive driving force for large recoveries. 

\section{Conclusion}
We have developed a protocol to predict ion separation through a 2D membrane highlighting the following core challenges: (i) ensuring correct charge distribution and removing capacitive effects when calculating ionic energy profiles in finite cells (ii) explicitly including first-principles solvation effects by machine-learning accelerated molecular dynamics (iii) contextualizing ion permeances within the larger picture of the electrochemical double layer structure and of the dynamical filtration process.

The results obtained on the crown-ether functionalized graphene membrane have illustrated and quantified key mechanisms for selective and efficient ion separation (notably, here, among isovalent ions), including chemically-driven individual ionic translocation barriers, electrostatic effects at the membrane interface, and steady-state charge separation upon filtration.      

It is noted that in the present model only the natural charge separation (Donnan effect) across the membrane has been considered. However, whenever conductive, the membrane may also be electrified thanks to an external voltage control, allowing to tune ion energy profiles and thus selectivities \cite{Patil2021,Leong2021}.    

Some methodology extensions to consider in the future are as follows. First, active learning can be used to make the neural network machine learning workflow more automatic and data-efficient, for instance via the fast uncertainty estimate approach recently introduced by Zhu et al. \cite{Zhu2023}. Second, including long-range electrostatic effects in the neural network potential\cite{Zhang2022} can allow for extended simulation sizes and thus capturing doubler layer effects beyond the mean-field approximation.     

\section{Methods}
\subsection{First-principles calculations}
Total energies are calculated by density-functional theory (DFT) with the Quantum ESPRESSO distribution \cite{QE}, using the Perdew-Burke-Ernzerhof (PBE) exchange correlation functional\cite{Perdew1996} in combination with pseudopotentials from the SSSP PBE efficiency 1.1.2 library for ionic cores\cite{SSSP}. Van der Waals interactions are included through Grimme's empirical correction DFT-D2 \cite{Grimme2012}. In the $2\times 2$ cell, the minimum energy is found by placing the ion at different positions above the pore and allowing it to relax in the transverse plane at the fixed $z$ value. 

\subsection{Capacitive correction scheme}
The extra electrostatic energy $\Delta E_{el}(z)$ of setup (ii) vs setup (i) is quantified as follows. Denoting by $\mathbf{D}_{1}$ and $\mathbf{D}_{2}$ the displacement fields in c.c. setups (i) and (ii), respectively, $\Delta E_{el}(z)$ is obtained, in atomic units (a.u.), as
\begin{equation}
8\pi \Delta E_{el}(z) = \int_{V}\mathbf{D}_{2}\mathbf{E}_{2}-\mathbf{D}_{1}\mathbf{E}_{1}dV = \int_{V}\frac{\mathbf{D}_{2}^{2}-\mathbf{D}_{1}^{2}}{\epsilon_{r}}dV
\label{eq:el_en}
\end{equation}
where $\epsilon_{r}$ is the relative permittivity -- arising mainly from the membrane electronic polarizability -- and $V$ the volume. If the ESM is placed far enough from the explicit system and does not modify the charge on the ion, then one has $\mathbf{D}_{2}=\mathbf{D}_{1}+\mathbf{D}_{u}$, where $\mathbf{D}_{u}$ is the uniform field $(0,0,\frac{2\pi e}{A})$ originating from the ESM, where $A$ denotes the transverse surface area of the simulation cell and $e$ the absolute charge of the electron. Rearranging Eq. \ref{eq:el_en}, and substituting $\mathbf{D}_{u}$, we have 

\begin{equation}
8\pi \Delta E_{el}(z)= \int_{V}\frac{2(\mathbf{D}_{2}-\mathbf{D}_{1})\mathbf{D}_{2}-(\mathbf{D}_{2}-\mathbf{D}_{1})^{2}}{\epsilon_{r}}dV = \int_{V}\frac{2\mathbf{D}_{u}\mathbf{D}_{2}-\mathbf{D}_{u}^{2}}{\epsilon_{r}}dV
\end{equation}

Finally, writing the electric field of setup (ii) as $\mathbf{D}_{2}/\epsilon_{r} = -\mathbf{\nabla} \Phi$, and dropping the second term in the numerator as it is independent from the ion position, we obtain, within a constant additive shift, 
\begin{equation}
\Delta E_{el}(z) = -\frac{2}{8\pi} \int_{V} \mathbf{D}_{u}\nabla \Phi dV = \frac{e\Phi_{D}(z)}{2}
\label{eqn:corr}
\end{equation}
where $\Phi_{D}(z)$ is the transversally averaged difference of the electrostatic potential $\Phi$ on the left-hand side vs the right-hand side of the simulation cell when the ion sits at $z$. Consequently, the EP of setup (ii) is corrected by subtracting the capacitive energy $\Delta E_{el}(z)$, as given by Eq. \ref{eqn:corr}, from the calculated total energy.

\begin{figure}\begin{center}
\includegraphics[width=0.6\columnwidth]{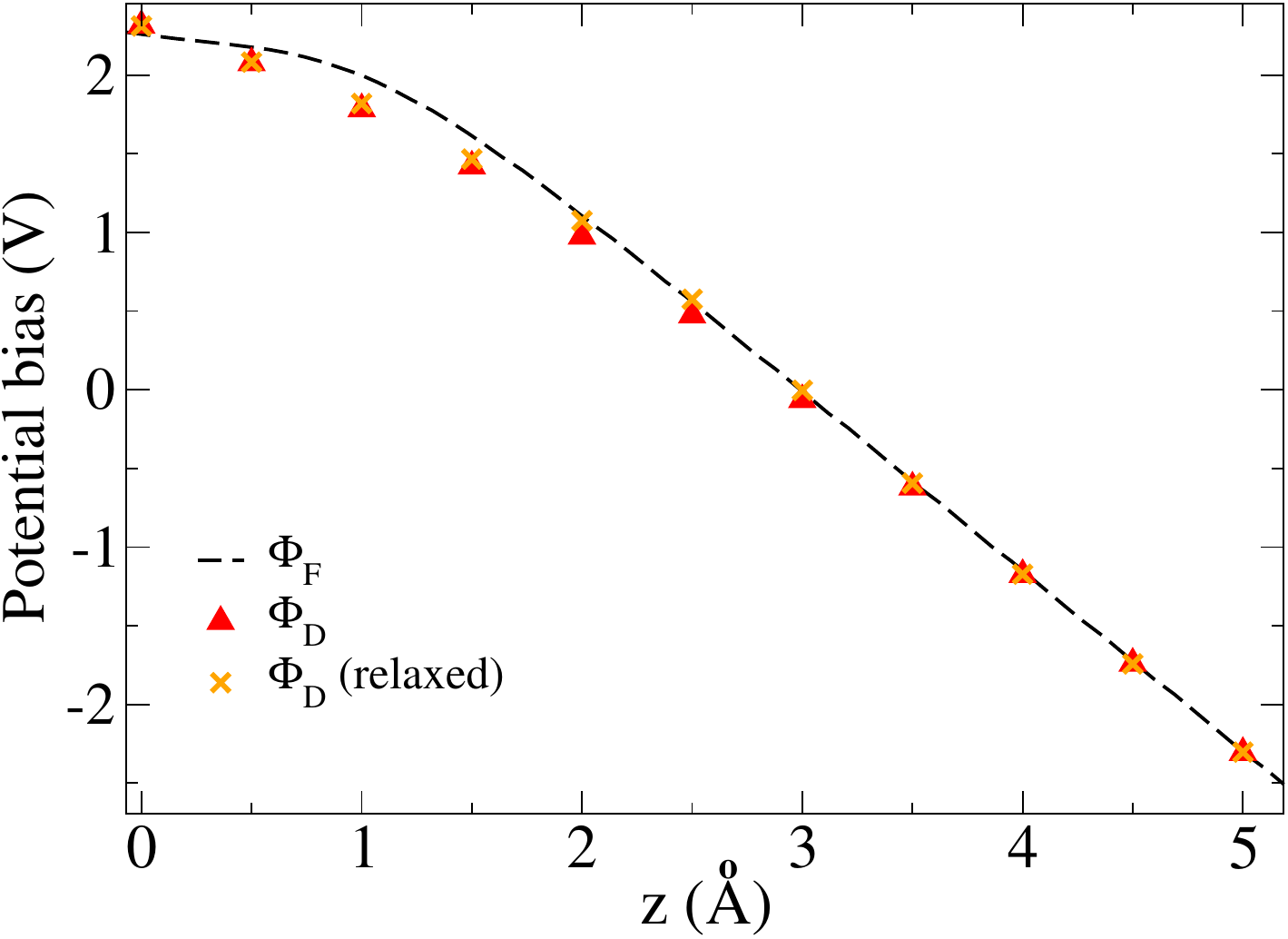}
\caption{Potential biases for different charging setups of equal intensity in the $1 \times 1$ cell: $\phi_{F}$ from applying an external electric field $\frac{4\pi e}{A}$ (a.u.) on the bare membrane ($z$ is then the generic longitudinal coordinate); $\Phi_{D}$ from adding Li$^{+}$ and its c.c. on the ESM ($z$ is then the ion position); and $\Phi_{D}$ (relaxed) when the membrane atomic positions are allowed to relax in response to the presence of Li$^{+}$ and its c.c.} 
\label{phi_graph}
\end{center}\end{figure}

For information, the potential bias $\Phi_{D}(z)$ is shown in Fig. \ref{phi_graph} for the unrelaxed and relaxed membrane. Alternatively, if one applies an external electric field of intensity $\frac{4\pi e}{A}$ to the bare membrane, the potential bias $\Phi_{F}(z)$ may be defined as the transversally averaged difference between the electrostatic potentials in the presence and absence of the field (here, exceptionally, $z$ is not the ion position but the generic longitudinal coordinate). In the present case, the fact that $\Phi_{D}(z) \approx \Phi_{F}(z)$ implies a negligible effect of the surface dipole generated by charge transfers between the membrane and the ion \cite{Schmickler2010,Bonnet2014}. 

\subsection{Machine-learning molecular dynamics}
The explicit solvent consists of 24 water molecules placed on the same side of the unrelaxed membrane as the ion, and confined within a potential wall placed at 7 {\AA} from the membrane to keep the average water density at 1 g/cm$^{3}$. The total energy function $E(z,\nu)$ is machine learned from a set of DFT calculations with the same functionals, pseudopotential library, and longitudinal OBC correction as in vacuum, and without a c.c. By contrast with vacuum calculations, no electronic charge spilling from the membrane to the ion is observed even at large distances.

The machine learning architecture is the E(3)-equivariant graph neural network (NN) Nequip of S. Batzner et al\cite{Batzner2022}. Three NN models (NN-1, NN-2, and NN-3) are used, with respective cutoff radii of the convolution filter of 5, 6, and 6 {\AA}, numbers of interaction blocks of 2, 3, and 3, and numbers of atomic features of 8, 8, and 32. The models are similar for all other parameters: a maximum rotation order of 2; no odd parity; a ``default" radial neural network comprising 8 basis radial functions, 3 layers, and 64 hidden neurons. Each NN model is trained over 100 epochs with the Adam optimizer, putting equal weights on forces and the total energy per atom in the cost function. The optimized models are then exported and used in LAMMPS\cite{Lammps} to perform MD simulations. The ML workflow consists of at least the following steps: (i) The NN-1 model is trained over 150 snapshots extracted from an initial first-principles MD simulation of a few ps; it is then used to run two MD simulations at 300 and 400 K over at least 50 ps; (ii) The NN-3 model is trained over 3000 snapshots extracted from the previous MD trajectories; it is then used to run MD simulations at 300 K (and optionally 400 K) over at least 50 ps. When necessary, the following steps were added: (iii) The NN-2 model is trained over 300 snapshots extracted from the 300 and 400 K MD simulations of steps (i) and (ii); it is then used to run two MD simulations at 300 and 400 K over at least 50 ps; (iv) The NN-3 model is re-trained over 3000 snapshots extracted from the previous MD trajectories; it is then used to run a MD simulation at 300 K. The energies predicted by the NN-3 MD trajectories are cross-validated with DFT calculations. The mean absolute error on the total energy is below 0.04 eV, in most cases after step (ii), and in some cases after step (iv). The final MD simulations are run with the corresponding NN-3 model to generate the EP in water. Simulation times greater than 500 ps are used to converge $E_{aq}^{1}(z)$ within 0.04 eV. 

\subsection{Microkinetic model of filtration}
\begin{figure}\begin{center}
\includegraphics[width=0.9\columnwidth]{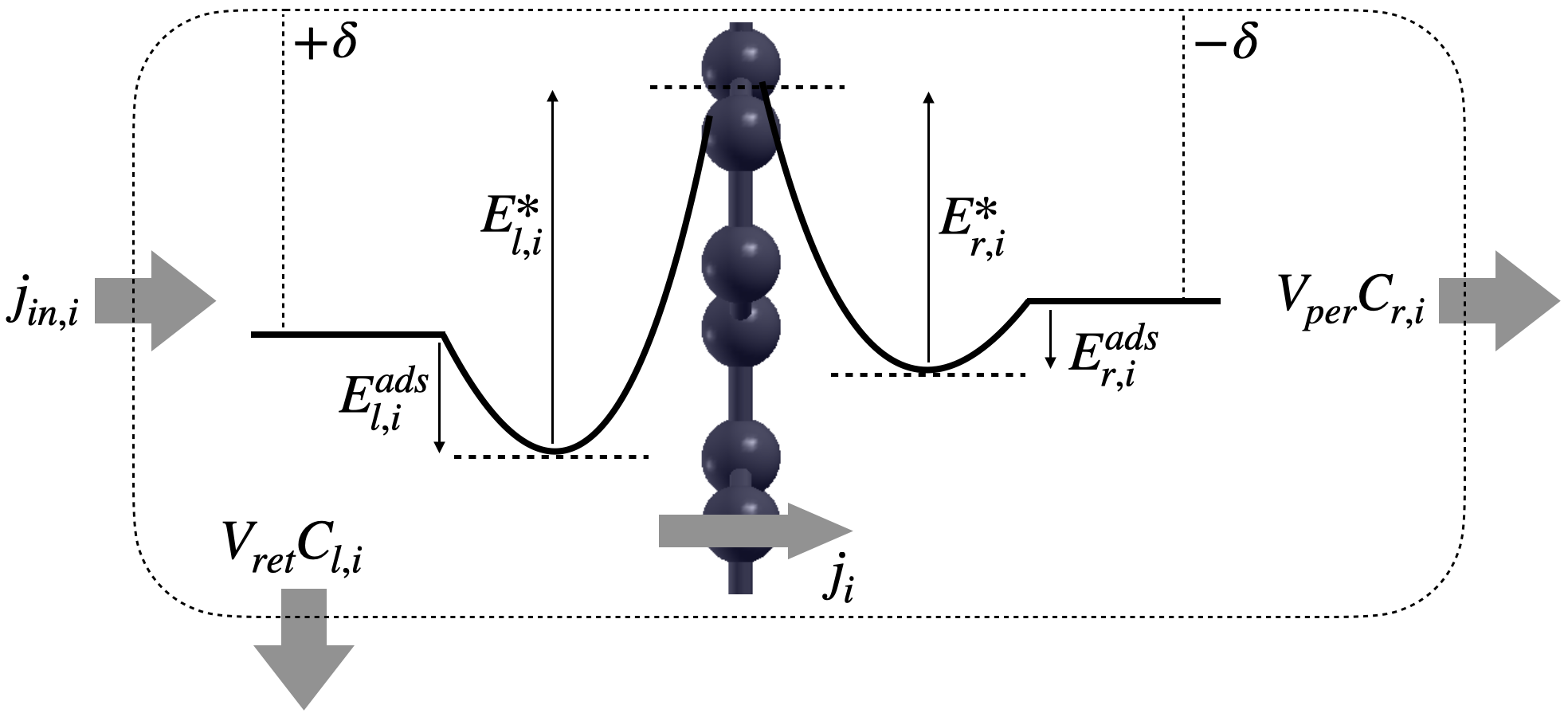}
\caption{Schematic representation of the filtration system.}
\label{physical_setup}
\end{center}\end{figure}

The filtration system is represented schematically in Fig. \ref{physical_setup}. The flow rates per membrane surface area are equivalent to velocities, denoted by $V_{ret}$ and $V_{per}$ for the retentate and permeate streams. The concentration of ion $i$ in each stream is respectively $C_{l,i}$ and $C_{r,i}$. A generic EP of ion $i$ is also illustrated, where we denote $E_{s,i}^{ads}$ the adsorption energy from the bulk solution onto the membrane pore, and $E_{s,i}^{*}$ the translocation barrier from the adsorption site to the pore center, with $s = l$ and $s = r$ for the left (retentate) and right (permeate) sides, respectively. Upon filtration, a surface charge separation density $\pm\delta$ is established across the membrane. For a given hydraulic regime, the system thus contains the $2 N_{ion}+1$ unknowns \{$C_{l,i}$, $C_{r,i}$, $\delta$\}, where $i \in \{ 1,...,N_{ion} \}$ and $N_{ion}$ is the number of ionic species. 

The steady state of the dynamical system is determined by the following set of $2 N_{ion}+1$ equations:
\begin{subequations} \label{eqn:line}
\begin{align}
j_{i}(C_{l,i},C_{r,i},\delta) & = V_{per}C_{r,i}                   \label{eqn:line1} \\
j_{in,i}                      & = V_{ret}C_{l,i} + V_{per}C_{r,i}  \label{eqn:line2} \\
\Sigma_{i} z_{i}C_{r,i}       & = 0                                \label{eqn:line3}
\end{align}
\end{subequations}
where Eq. \ref{eqn:line1} and \ref{eqn:line2} express the conservation of molar fluxes, and Eq. \ref{eqn:line3} is the electroneutrality condition ($z_{i}$ denoting the valency of ion $i$). In Eq. \ref{eqn:line1}, $j_{i}$ is the net molar flux of ion $i$ across the membrane, and in Eq. \ref{eqn:line2}, $j_{in,i}$ is its incoming molar flux upstream of the membrane. The molar flux $j_{i}$ is expressed by the transition-state theory (TST)\cite{Chorkendorff2007} as 

\begin{equation}
j_{i} = j_{i}^{f} - j_{i}^{b} = j_{0} \left[ \theta_{l,i}(1-\theta_{r,i})e^{-\frac{E_{l,i}^{*}}{kT}} - \theta_{r,i}(1-\theta_{l,i})e^{-\frac{E_{r,i}^{*}}{kT}} \right]
\label{eqn:current}
\end{equation}
with $j_{0} = \frac{\alpha_{p} kT}{h}$, where $j_{i}^{f}$ and $j_{i}^{b}$ are the forward and backward fluxes across the membrane, $\alpha_{p}$ the pore surface density, and $\theta_{s,i}$ the adsorption site average occupation (between 0 and 1) by ion $i$ on side $s$. In turn, $\theta_{s,i}$ is determined by the steady-state condition,

\begin{equation}
j_{0}(1-\theta_{s,i}) C_{s,i} e^{-\frac{s_{b,i}}{k} } - \left(j_{0}\theta_{s,i} e^{ \frac{E_{s,i}^{ads}}{kT} } + \zeta_{s} j_{i} \right) = 0
\label{eqn:theta}
\end{equation}

expressing the balance between incoming (first term) and outgoing (second term) ions at the adsorption site, with $\zeta_{s}=$ +1 and $-1$ for the left and right sides, respectively. Here, $s_{b,i}$ is the entropy of ion $i$ at a unit concentration in the bulk and is approximated using the gas-phase translational entropy formula\cite{Chorkendorff2007}, $s_{b,i} = k \left[ \frac{3}{2} + ln \left( \Omega\left( \frac{2\pi m_i kT}{h^{2}} \right)^{\frac{3}{2}} \right)\right]$, where $m_i$ is the mass of the ion, and $\Omega$ the volume per ion at the specified concentration. The use of the bulk concentration $C_{s,i}$ in the first term further assumes that mass-transfer mechanisms from the bulk solution to the membrane interface are not rate-limiting. The dependence of $E_{s,i}^{ads}$ and $E_{s,i}^{*}$ on the electrostatic environment in the vicinity of the membrane is expressed in a mean-field fashion by the following equations:  

\begin{subequations} \label{eqn:lline}
\begin{align}
E_{l,i}^{ads}(\delta,\theta_{l}) & =  E_{i}^{ads} + ez_{i}\Phi_{DL}(z_{ads},-\delta+\sigma_{i}(\theta_{l,i})) \label{eqn:lline1} \\
E_{l,i}^{*}(\delta)              & =  E_{i}^{*}   + ez_{i}\left[ \Phi_{DL}(0,-\delta) - \Phi_{DL}(z_{ads},-\delta) \right] \label{eqn:lline2} \\
E_{r,i}^{ads}(\delta,\theta_{r}) & =  E_{i}^{ads} + ez_{i}\Phi_{DL}(z_{ads},\delta+\sigma_{i}(\theta_{r,i})) \label{eqn:lline3} \\
E_{r,i}^{*}(\delta)              & =  E_{i}^{*}   + ez_{i}\left[ \Phi_{DL}(0,\delta) - \Phi_{DL}(z_{ads},\delta) \right]   \label{eqn:lline4}                     
\end{align} 
\end{subequations}

where $E_{i}^{ads}$ and $E_{i}^{*}$ are the energies at zero surface charge as obtained from the first-principles solvated EPs; $\sigma_{i}(\theta) = \alpha_{p}ez_{i}\theta$ is the surface charge density created by the ion adsorbates; and $\Phi_{DL}(z,\sigma_{tot})$ is the EDL potential bias at a distance $z$ from the membrane carrying a total effective surface charge density of $\sigma_{tot}$, with the convention $\Phi_{DL}(+\infty,\sigma_{tot}) = 0$. Induced surface dipole effects are neglected, hence the use of $E_{DL}(z)$ = $ez_{i}\Phi_{DL}(z,\sigma_{tot})$ as the ion chemical potential shift at $z$ upon EDL charging. The rationale behind Eq. \ref{eqn:lline1} is that an ion approaching the adsorption site (at position $z_{ads}$) from the left bulk solution faces a net charge of $-\delta +\sigma_{i}(\theta_{l,i})$ distributed behind and in front of the membrane. Beyond this point, however, only the charge $-\delta$ is felt by the ion as expressed by Eq. \ref{eqn:lline2}. The right-hand side energy dependences are equivalently given by Eqs \ref{eqn:lline3} and \ref{eqn:lline4}. It is noted that although all ionic species in principle compete for the same adsorption sites, here for simplicity this competition is neglected in Eqs \ref{eqn:theta} and \ref{eqn:lline}. However, if the different ionic species have roughly similar adsorption energies (as in the present case study), then this decoupling assumption provides a reasonable approximation.   

In general, the function $\Phi_{DL}(z,\sigma_{tot})$ depends on the complex structure of the electrochemical double layer including at least the contributions of the Stern and Gouy-Chapman layers \cite{Shin2022}. At sufficiently high concentrations ($>$ 0.01 M, brackish water), however, the potential bias is dominated by the Stern layer. Furthermore, the Stern potential difference occurs mainly within the mostly dehydrated and thus low-permittivity $0 < z < 3$ region. Consequently, we can use the approximation $\Phi_{DL}(z,\sigma_{tot}) \approx \sigma_{tot} \Phi(z)\left[ 1-H(z-3) \right]$, where $\Phi(z)$ is the potential bias profile determined previously (Fig. \ref{phi_graph}) renormalized to a unit surface charge density, and $H(z)$ is the Heaviside function.    

For given values of $C_{l,i}$, $C_{r,i}$, and $\delta$, the $\theta_{s,i}$ are found by dichotomy through Eqs \ref{eqn:theta} and \ref{eqn:lline}, and $j_{i}$ is inferred through Eqs \ref{eqn:current} and \ref{eqn:lline}. Then, for a fixed $\delta$ the ionic concentrations are determined through Eqs \ref{eqn:line1} and \ref{eqn:line2}, and finally $\delta$ is determined through Eq. \ref{eqn:line3}.

It should be noted that although $V_{per}$ has been presented as a permeate stream velocity as per customary filtration terminology, the model is actually agnostic to the provenance of the stream, which can consist partially or totally of fresh make-up water on the downstream side. Correspondingly, the ionic translocation flux expressed in Eq. \ref{eqn:current} does not rely on an advective water flux across the membrane. 

\begin{figure}\begin{center}
\includegraphics[width=0.5\columnwidth]{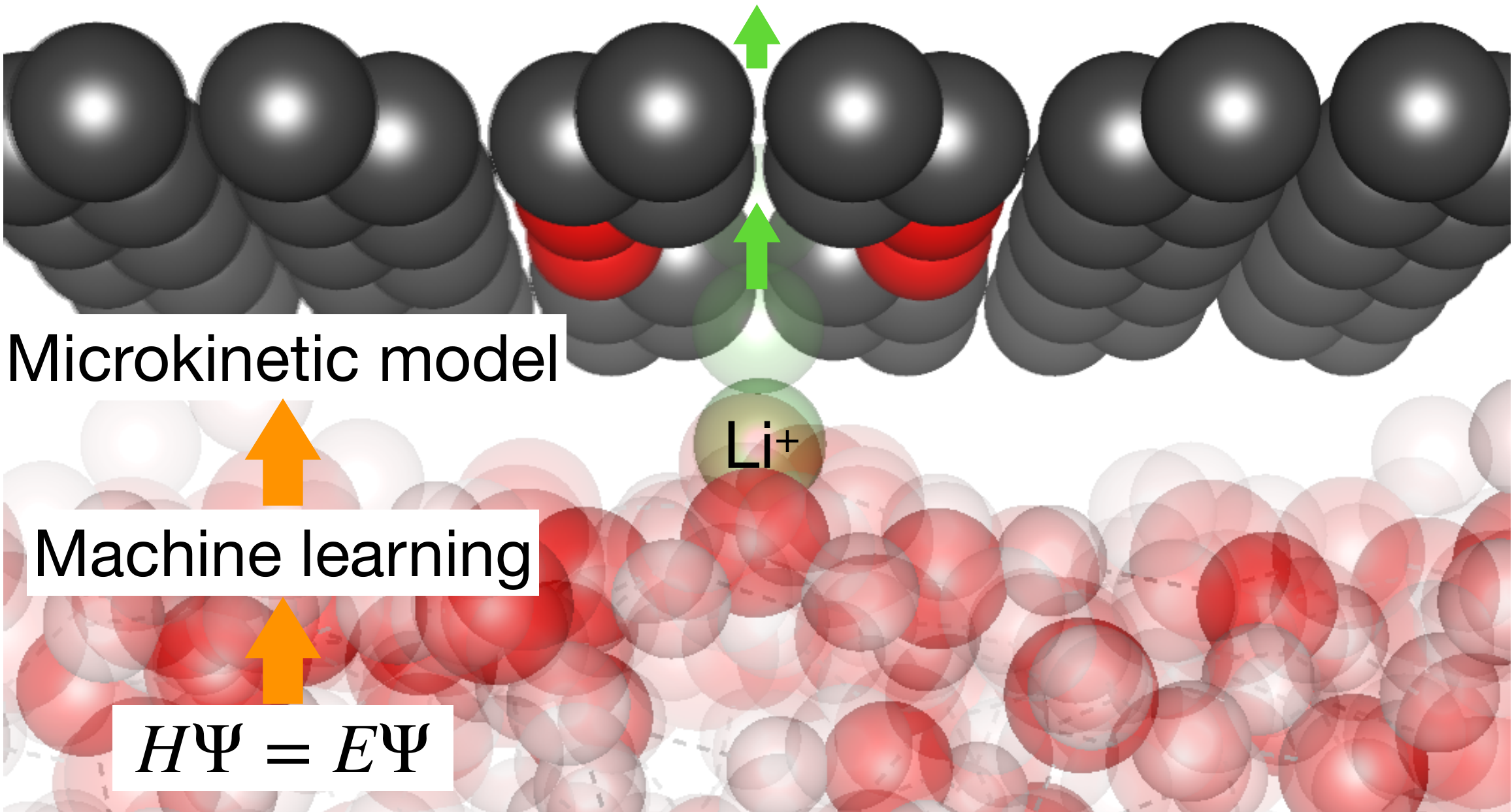}
\caption{Graphical Table of Content.}
\end{center}\end{figure}

\section{Data availability}
MD trajectories, energies and forces used to obtain the energy and free energy profiles of Fig. \ref{all_pes} are available on Materials Cloud (DOI: 10.24435/materialscloud:mg-wh).

\section{Acknowledgements}
N. B. acknowledges funding from the European Union’s Horizon 2020 research and innovation programme under the Marie Skłodowska-Curie grant agreement No 101034260. This work was supported by the Swiss National Supercomputing Centre (CSCS) grant under
project ID s1192.


\begin{thebibliography}{00}

\bibitem{Shao2020}
  Shao, L. Grand challenges in emerging separation technologies. \textit{Front. Environ. Chem.} \textbf{2020}, \textit{1}, 3-1--5.

\bibitem{DuChanois2021}
  DuChanois, R.M.; Porter, C.J.; Violet, C.; Verduzco, R.; Elimelech, M. Membrane materials for selective ion separations at the water-energy nexus. \textit{Adv. Mater.} \textbf{2021}, \textit{33}, 2101312-1--18.
  
\bibitem{Werber2016}
  Werber, J.R.; Deshmukh, A.; Elimelech, M. The critical need for increased selectivity, not increased water permeability, for desalination membranes. \textit{Environ. Sci. Technol. Lett.} \textbf{2016}, \textit{3}, 112--120.  
  
\bibitem{Stringfellow2021}
  Stringfellow, W.T.; Dobson, P.F. Technology for the recovery of lithium from geothermal brines. \textit{Energies} \textbf{2021}, \textit{14}, 6805-1--72. 
  
  \bibitem{Mounet2018}
  Mounet, N.; Gibertini, M.; Schwaller, P.; Campi, D.; Merkys, A.; Marrazzo, A.; Sohier, T.; Castelli, I.E.; Cepellotti, A.; Pizzi, G.; Marzari, N. Two-dimensional materials from high-throughput computational exfoliation of experimentally known compounds. \textit{Nat. Nanotech.} \textbf{2018}, \textit{13}, 246--252.  
  
 \bibitem{Zhao2024}
  Zhao, K.; Lee, W.C.; Rezaei, M.; Chi, H.Y.; Li, S.; Villalobos, L.F.; Hsu, K.J.; Zhang, Y.; Wang, F.C.; Agrawal, K.V. Tuning pore size in graphene in the angstrom regime for highly selective ion-ion separation. \textit{ACS Nano} \textbf{2024}, \textit{18}, 5571--5580.

 \bibitem{Szymczyk2005}
  Szymczyk, A.; Fievet, P. Investigating transport properties of nanofiltration membranes by means of steric, electric and dielectric exclusion model. \textit{J. Memb. Sci.} \textbf{2005}, \textit{252}, 77--88.
  
  \bibitem{Yuan2022}
  Yuan, Z.; He, G.; Xin Li, S.; Misra, R.P.; Strano, M.S.; Blankschtein, D. Gas separations using nanoporous atomically thin membranes: recent theoretical, simulation, and experimental advances. \textit{Adv. Mater.} \textbf{2022}, \textit{34}, 2201472-1--22.
 
 \bibitem{Wen2016}
  Wen, Q.; Yan, D.; Liu, F.; Wang, M.; Ling, Y.; Wang, P.; Kluth, P.; Schauries, D.; Trautmann, C.; Apel, P.; Guo, W; Xiao, G.; Liu, J.; Xue, J.; Wang, Y. Highly selective ionic transport through subnanometer pores in polymer films. \textit{Adv. Funct. Mater.} \textbf{2016}, \textit{26}, 5796--5803.
 
 \bibitem{Razmjou2019}
  Razmjou, A.; Asadnia, M.; Hosseini, E.; Korayem, A.H.; Chen, V. Design principles of ion selective nanostructured membranes for the extraction of lithium ions. \textit{Nat. Comm.} \textbf{2019}, \textit{10}, 5793-1--15. 
  
  \bibitem{Fan2023}
  Fan, H.; Huang, Y.; Yip, N.Y. Advancing ion-exchange membranes to ion-selective membranes: principles, status, and opportunities. \textit{Front. Environ. Sci. Eng.} \textbf{2023}, \textit{17}, 25-1--27.
 
 \bibitem{Fievet2014}
  Fievet, P. \textit{Encyclopedia of Membranes -- Donnan effect}; Springer, 2014.

\bibitem{Fang2019}
  Fang, A.; Kroenlein, K.; Riccardi, D.; Smolyanitsky, A. Highly mechanosensitive ion channels from graphene-embedded crown ethers. \textit{Nat. Mater.} \textbf{2019}, \textit{18}, 76--81. 
  
\bibitem{Sahu2019}
  Sahu, S.; Elenewski, J.; Rohmann, C.; Zwolak, M. Optimal transport and colossal ionic mechano-conductance in graphene crown ethers. \textit{Sci. Avd.} \textbf{2019}, \textit{5}, eaaw5478-1--7.
  
\bibitem{Guo2021}
  Guo, K.; Liu, S.; Tu, H.; Wang, Z.; Chen, L.; Lin, H.; Miao, M; Xu, J.; Liu, W. Crown ethers in hydrogenated graphene. \textit{Phys. Chem. Chem. Phys.} \textbf{2021}, \textit{23}, 18983--18989.
  
\bibitem{Lv2023}
  Lv, Y.; Dong, L.; Cheng, L.; Gao, T.; Wu, C.; Chen, X.; He, T; Cui, Y.; Liu, W. Tailoring monovalent ion sieving in graphene-oxide membranes with high flux by rationally intercalating crown ethers. \textit{ACS Appl. Mater. Interfaces} \textbf{2023}, \textit{15}, 46261--46268.
  
\bibitem{Ritt2022}
  Ritt, C.L.; Liu, M.; Pham, T.A.; Epstein, R.; Kulik, H.J.; Elimelech M. Machine learning reveals key ion selectivity mechanisms in polymeric membranes with subnanometer pores. \textit{Sci. Adv.} \textbf{2022}, \textit{8}, eabl5771-1--9. 
  
  \bibitem{Zhou2020}
  Zhou, X.; Wang, Z.; Epsztein, R.; Zhang, C.; Li, W.; Fortner, J.D.; Pham, T.A.; Kim, J.H.; Elimelech, M. Intrapore energy barriers govern ion transport and selectivity of desalination membranes. \textit{Sci. Adv.} \textbf{2020}, \textit{6}, eabd9045-1--9. 
  
   \bibitem{Chipot2023}
  Chipot, C. Predictions from first-principles of membrane permeability to small molecules: How useful are they in practice. \textit{J. Chem. Inf. Model.} \textbf{2023}, \textit{63}, 4533--4544. 
  
\bibitem{Sint2008}
  Sint, K.; Wang, B.; Kral, P. Selective ion passage through functionalized graphene nanopores. \textit{J. Am. Chem. Soc.} \textbf{2008}, \textit{130}, 16448--16449. 
  
\bibitem{Suk2010}
  Suk, M.E.; Aluru, N.R. Water transport through ultrathin graphene. \textit{Phys. Chem. Lett.} \textbf{2010}, \textit{1}, 1590--1594. 
  
\bibitem{Ruan2016}
  Ruan, Y.; Zhu, Y.; Zhang, Y.; Gao, Q.; Lu, X.; Lu, L. Molecular dynamics study of Mg$^{2+}$/Li$^{+}$ separation via biomimetic graphene-based nanopores: The role of dehydration in second shell. \textit{Langmuir} \textbf{2016}, \textit{32}, 13778--13786.
  
\bibitem{Smolyanitsky2018}
  Smolyanitsky, A.; Paulechka, E.; Kroenlein, K. Aqueous ion trapping and transport in graphene-embedded 18-crown-6 ether pores. \textit{ACS Nano} \textbf{2018}, \textit{12}, 6677--6684.  
  
\bibitem{Zofchak2022}
  Zofchak, E.S.; Zhang, Z.; Marioni, N.; Duncan, T.J.; Sachar, H.S.; Chamseddine, A.; Freeman, B.D.; Ganesan, V. Cation-ligand interactions dictate salt partitioning and diffusivity in ligand-functionalized polymer membranes. \textit{Macromolecules} \textbf{2022}, \textit{55}, 2260--2270. 
  
\bibitem{Meng2023}
  Meng, K.; Zhao, X.; Niu, Y.; Ming, S.; Xu, J.; Hou, H.; Yu, X.; Rong, J. Understanding the desalination mechanism of a two-dimensional graphene-like membrane using data-driven design. \textit{Diam. Rel. Mater.} \textbf{2023}, \textit{137}, 110085-1--10.    

\bibitem{Andreussi2012}
Andreussi, O.; Dabo, I.; Marzari, N. Revised self-consistent continuum solvation in electronic-structure calculations. \textit{J. Chem. Phys.} \textbf{2012}, \textit{136}, 064102-1--20.    
   
\bibitem{Shin2022}
  Shin, S.J.; Kim, D.H.; Bae, G.; Ringe, S.; Choi, H.; Lim, H.K.; Choi, C.H.; Kim, H. On the importance of the electric double layer structure in aqueous electrocatalysis. \textit{Nat. Comm.} \textbf{2022}, \textit{13}, 174-1--8.   
  
  \bibitem{Chorkendorff2007}
  Chorkendorff, I.; Niemantsverdriet, J.W. \textit{Concepts of Modern Catalysis and Kinetics}; Wiley, 2007.  
   
  \bibitem{Otani2006}  
Otani, M.; Sugino, O. First-principles calculations of charged surfaces and interfaces: A plane-wave nonrepeated slab approach. \textit{Phys. Rev. B} \textbf{2006}, \textit{73}, 115407-1--11.  
   
 \bibitem{Environ}
http://www.quantum-environ.org/ (accessed July 25, 2024). 
   
\bibitem{Dabo2008}
  Dabo, I.; Kozinsky, B.; Singh-Miller, N.; Marzari, N. Electrostatics in periodic boundary conditions and real-space corrections \textit{Phys. Rev. B} \textbf{2008}, \textit{77}, 115139-1--13.   
   
   \bibitem{Chan2015}
  Chan, K.; Norskov, J.K. Electrochemical barriers made simple. \textit{J. Phys. Chem. Lett.} \textbf{2015}, \textit{6}, 2663--2668. 
  
  \bibitem{Tang2009}
 Tang, W.; Sanville, E.; Henkelman, G. A grid-based Bader analysis algorithm without lattice bias \textit{J. Phys. Condens. Matter} \textbf{2009}, \textit{21}, 084204-1--7. 
 
 \bibitem{Milman1993}
Milman, V.; Payne, M.C.; Heine, V.; Needs, R.J.; Lin, J.S.; Lee, M.H.. Free energy and entropy of diffusion by ab initio molecular dynamics: Alkali ions in silicon. \textit{Phys. Rev. Lett.} \textbf{19993}, \textit{70}, 2928--2931.
 
 \bibitem{Conway1978}
Conway, B. E. The evaluation and use of properties of individual ions in solution. \textit{J. Solution Chem.} \textbf{1978}, \textit{10}, 721--770.
 
  \bibitem{Xi2022}
  Xi, C.; Zheng, F.; Gao, G.; Song, Z.; Zhang, B.; Dong, C.; Du, X.W.; Wang, L.W. Ion solvation free energy calculation based on ab initio molecular dynamics using a hybrid solvent model \textit{J. Chem. Theory Comput.} \textbf{2022}, \textit{18}, 6878-6891.
  
  \bibitem{Patil2021}
Patil, J.J.; Jana, A.; Getachew, B.A.; Bergsman, D.S.; Gariepy, Z.; Smith, B.D.; Lu, Z.; Grossman, J.C. Conductive carbonaceous membranes: recent progress and future opportunities. \textit{J. Mater. Chem. A} \textbf{2021}, \textit{9}, 3270--3289. 

\bibitem{Leong2021}
Leong, Z.Y.; Han, Z.; Wang, G.; Li, D.S.; Yang, S.A.; Yang, H.Y. Electric field modulated ion-sieving effects of graphene oxide membranes. \textit{J. Mater. Chem. A} \textbf{2021}, \textit{9}, 244--253. 

\bibitem{Zhu2023}
Zhu, A.; Batzner, S.; Musaelian, A.; Kozinsky, B. Fast uncertainty estimates in deep learning interatomic potentials. \textit{J. Chem. Phys.} \textbf{2023}, \textit{158}, 164111-1--9.

\bibitem{Zhang2022}
Zhang, L.; Wang, H.; Muniz, M.C.; Panagiotopoulos, A.Z.; Car, R.; E, W. A deep potential model with long-range electrostatic interactions. \textit{J. Chem. Phys.} \textbf{2022}, \textit{156}, 124107-1--14.
   
 \bibitem{QE}
https://www.quantum-espresso.org/ (accessed July 25, 2024).
   
\bibitem{Perdew1996}  
Perdew, J. P.; Burke, K.; Ernzerhof, M. Generalized gradient approximation made simple. \textit{Phys. Rev. Lett.} \textbf{1996}, \textit{77}, 3865--3868.   
   
  \bibitem{SSSP}
 https://www.materialscloud.org/discover/sssp/table/efficiency (accessed July 25, 2024).
   
  \bibitem{Grimme2012}  
Grimme, S. Supramolecular binding thermodynamics by dispersion-corrected density functional theory. \textit{Chem. Eur. J.} \textbf{2012}, \textit{18}, 9955--9964. 
  
\bibitem{Schmickler2010}
  Schmickler, W.; Santos, E. \textit{Interfacial Electrochemistry}; Springer, 2010.
  
 \bibitem{Bonnet2014}
 Bonnet, N.; Dabo, I.; Marzari, N. Chemisorbed molecules under potential bias: Detailed insights from first-principles vibrational spectroscopies \textit{Electrochimica Acta} \textbf{2014}, \textit{121}, 210--214.  
 
           
 \bibitem{Batzner2022}
 Batzner, S.; Musaelian, A.; Sun, L.; Geiger, M.; Mailoa, J.P.; Kornbluth, M.; Molinari, N.; Smidt, T.E.; Kozinsky, B. E(3)-equivariant graph neural networks for data-efficient and accurate interatomic potentials \textit{Nat. Comm.} \textbf{2022}, \textit{13}, 2453-1--11.  

 \bibitem{Lammps}
 https://www.lammps.org/ (accessed July 25, 2024).    

 
\end{thebibliography}
\end{document}